\documentclass[conference]{IEEEtran}
\IEEEoverridecommandlockouts
\usepackage{amsmath,amssymb,amsfonts}
\usepackage{algorithmic}
\usepackage{graphicx}
\usepackage{textcomp}
\usepackage{xcolor}
\usepackage{biblatex}
\bibliography{ExportedItems}
\def\BibTeX{{\rm B\kern-.05em{\sc i\kern-.025em b}\kern-.08em
    T\kern-.1667em\lower.7ex\hbox{E}\kern-.125emX}}
\begin{document}

\title{Smartphone Vibrometric Force Estimation for Grip Related Strength
Measurements\\
\thanks{Identify applicable funding agency here. If none, delete this.}
}

\author{\IEEEauthorblockN{Colin Barry}
\IEEEauthorblockA{\textit{Electrical and Computer Engineering} \\
\textit{University of California San Diego}\\
San Diego, USA \\
c1barry@ucsd.edu}
\and
\IEEEauthorblockN{Byron Fergerson}
\IEEEauthorblockA{\textit{Department of Anesthesiology} \\
\textit{UC San Diego School of Medicine}\\
San Diego, USA }
\and
\IEEEauthorblockN{Edward Jay Wang}
\IEEEauthorblockA{\textit{Electrical and Computer Engineering} \\
\textit{University of California San Diego}\\
San Diego, USA }

}

\maketitle

\begin{abstract}
Hand grip strength (HGS) is a widely used clinical biomarker linked to mobility, frailty, surgical outcomes, and overall health. This work explores a novel, phone-only approach for estimating grip-related force using a smartphone’s built-in vibration motor and inertial measurement unit. When the phone vibrates, applied finger force modulates the amplitude of high-frequency accelerometer and gyroscope signals through Vibrometric Force Estimation (VFE). We profiled a Google Pixel 4 using synchronized IMU data and ground-truth force measurements across varied force trajectories, then trained ridge-regression models for both absolute and relative force prediction. In 15-fold hold-one-out validation, absolute force estimation achieved a mean absolute error of 1.88 lb, while relative force estimation achieved a mean error of 10.1\%. Although the method captures pinch-type force rather than standardized full-hand HGS, the results demonstrate the feasibility of smartphone-based strength assessment using only on-device sensors. This approach may enable large-scale, low-burden functional health measurements once profiling is completed for major smartphone models.
\end{abstract}


\section{Introduction}
This paper explores the feasibility of a novel sensing solution using a smartphone’s built-in vibration motor and accelerometer to enable proxy HGS measurements. Hand grip strength (HGS) measurements can serve as important biomarkers especially for older adults \cite{bohannon_grip_2019} because HGS is highly correlated with all cause mortality\cite{chai_comparison_2024}, surgical morbidity \cite{griffith_delayed_2005}, bone mineral density \cite{song_causal_2022}, fracture risk \cite{song_causal_2022, us_preventive_services_task_force_screening_2025}, old age disability \cite{rantanen_midlife_1999}, muscle mass \cite{mcgrath_handgrip_2018, meza-valderrama_sarcopenia_2021}, and even cognitive impairments \cite{huang_association_2022}. For this reason, HGS is an essential component of all sarcopenia screening methods \cite{kirk_sarcopenia_2021, mayhew_sarcopenia_2023, cruz-jentoft_sarcopenia_2010, meza-valderrama_sarcopenia_2021}. This novel solution will be paired with known physical performance tasks that can be implemented on the smartphone to provide a multifaceted approach to sarcopenia monitoring, surgical readiness screening, rehabilitation assessments, and general fitness \cite{cruz-jentoft_sarcopenia_2010, kirk_sarcopenia_2021, myles_how_2024, rijk_prognostic_2016, sousa-santos_differences_2017}.

The key to capturing the HGS of a person is a clever technological insight called Vibrometric Force Estimation (VFE). VFE is a technique that leverages the natural damping effects when a force is applied to a vibrating object, allowing the inference of the forces that are applied to a smartphone. While vibrating the motor on a smartphone, a user grips the phone with their hand, similar to a clinical hand grip dynamometer. The smartphone’s accelerometer measures the amplitude of vibration which is damped in relation to the force amplitude induced by the grip. This measurement provides the basis of our proposed smartphone dynamometer system. 

With nearly all modern smartphones containing an integrated vibration motor and inertial measurement unit (IMU), the proposed smartphone HGS measurement system is highly feasible for widespread use. A smartphone IMU contains an accelerometer, a gyroscope, and possibly other sensors to capture movement. Although these components exist in nearly every smartphone, smartphone models differ in shape, size, materials, specific sensors, and component locations. For this reason, a given smartphone model must be profiled in order to allow for accurate measurements of applied force onto the smartphone. The profiling involves recording from the accelerometer and gyroscope of the smartphone IMU while applying a known force. This data is then used to train a model that converts filtered accelerometer and gyroscope data to applied force.

This work demonstrates the feasibility of a proxy HGS measurement on a smartphone. The measurement tested in this paper is a measurement of applied force between the index finger and the thumb as placed on either side of the smartphone. This metric differs from the standardized HGS metric on hand dynamometers and is more akin to a pinch gauge metric. This altered method is chosen to simplify the positioning of the hand on the phone and reduce variability in the way an individual can apply force. The future vision will be to profile a set of the most popular smartphones to proliferate this measurement and enable widespread use of proxy HGS measurements.

\begin{figure*}
    \centering
    \includegraphics[width=10cm]{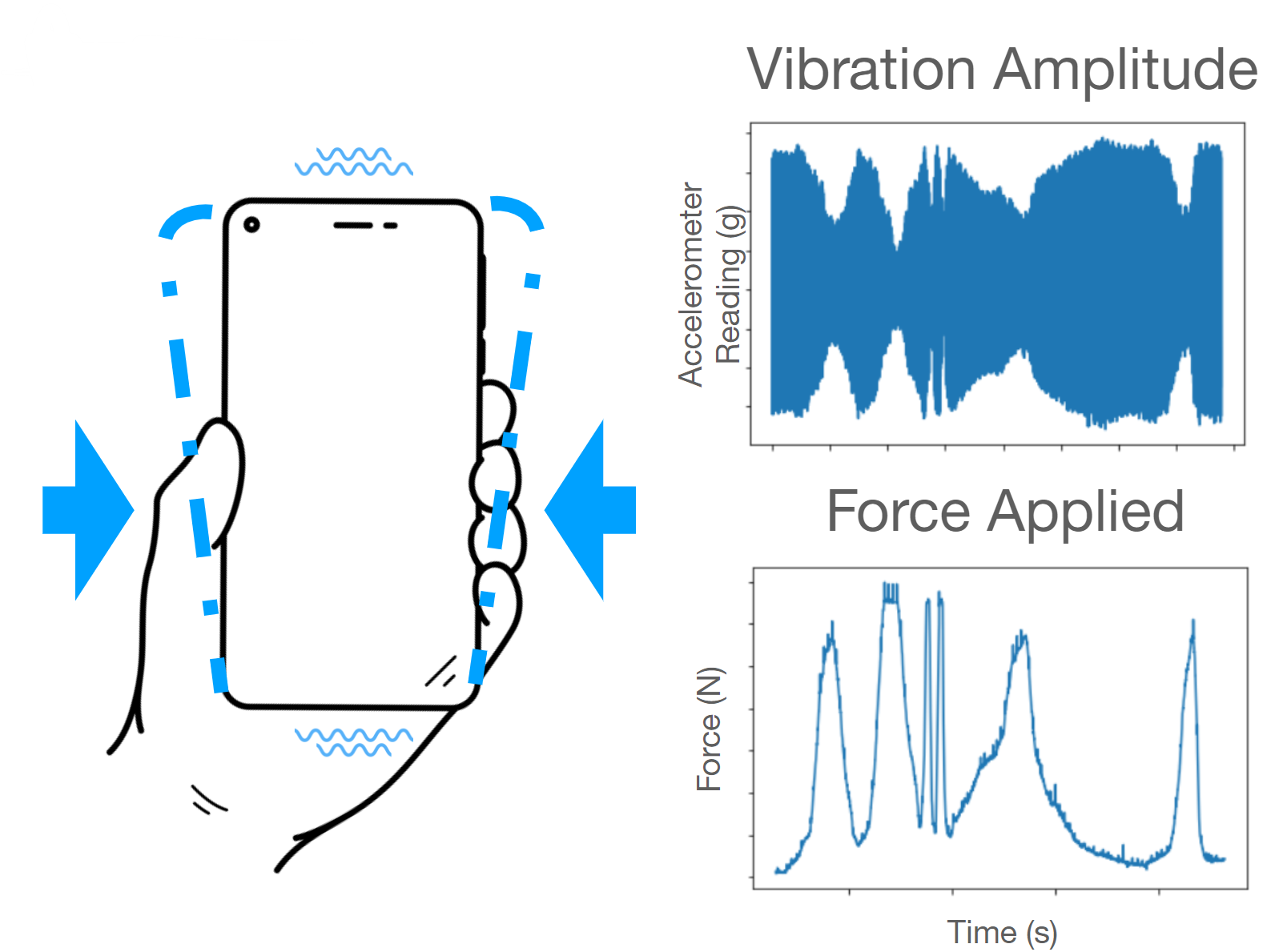}
    \caption{Overview of Smartphone Hang Grip Strength Measurement. Part a) A hand squeezing a smartphone while the phone is vibrating. Part b) Smartphone accelerometer signal amplitude during vibration correlates with applied force intensity.}
    \label{fig:overview}
\end{figure*}

\subsection{Related Work}
Though this work represents the first smartphone proxy for HGS measurement, several works have explored utilizing the vibration motor and IMU for sensing. The most notable prior work is the author's prior work on smartphone blood pressure measurements via the oscillometric finger pressing method\cite{barry_oscillometric_2024}. This work involves VFE to determine the applied force of the finger. The blood pressure metric requires positioning a smartphone flat on a table with the user applying light force (approximately 1-3 lbs) to create a signal at the fingertip similar to that of a blood pressure cuff at the upper arm. The proposed work significantly differs because the user must apply force between their thumb and index finger, applying force likely more than 10 times that of the oscillometric finger pressing method. The difference in amplitude and positioning significantly changes the force modeling, end application, and usability.

Other prior work includes utilizing the vibration motor and gyroscope to detect hand positioning on the smartphone, including levels of grip. This work focused on sensing hand positioning as an element to be considered in designs for User Interface and User Experience (UI/UX)\cite{goel_gripsense_2012}. This work did not explore a continuous range of force measurements for the application of a HGS measurement.

Another prior work investigated the potential to use the smartphone as a scale. This measurement involved placing objects on top of the smartphone while the smartphone vibrated to estimate the weight of the object by the change in vibration\cite{zhang_vibroscale_2020}. This conference poster involved a limited analysis compared to full papers, and did not consider HGS or other functional measurements.

\section{Methods}\label{sec11}

\subsection{Force Estimation Profiling}
The proposed smartphone measurement method is largely dependent on accurately estimating force damping without any attachments. Each smartphone model has its own force dampening effect profile. To characterize the profile, a small-scale data collection is performed on each phone model to capture the vibration modulation at different known forces measured by force sensitive resistors. 

We conducted a thorough profiling on a smartphone Google Pixel 4. The smartphone model profiling involves applying force on the phone with a 0.3mm thick force sensor (SingleTact 15mm 100N) positioned under the index finger. The user positions their thumb and index finger at indicated positions on the smartphone and applies a range of pressure. The force sensor under the index finger measures the applied force while the phone simultaneously records from the IMU. The IMU data and force sensor data are later aligned to create a dataset that will be used to train a force model that can predict applied force from IMU data.

During a measurement, real-time readings from the ground truth force sensor are displayed on a computer screen in front of the user with a force guide. The user is instructed to use real time feedback to follow the force guide in applying a range of pressures. The study involves 40 second sessions where a participant applies a range of forces. The forces are varied for different sessions, where some sessions involve linear increases, while others involve patterns like sine waves or square waves.

\subsection{Data Analysis}
\subsubsection{Signal Processing}
Raw accelerometer and gyroscope signals were preprocessed to extract features relevant for force prediction. First, signals were band-pass filtered using a 4th-order Butterworth filter centered at 136 Hz (the dominant frequency of the vibration) with a 10 Hz bandwidth. The full envelope of each filtered signal was then computed  as the sum of the upper envelope and the absolute value of the lower envelope.
\subsubsection{Temporal Alignment}
To account for temporal misalignment between IMU signals and force measurements, cross-correlation analysis was performed to determine optimal lag offsets. For each measurement, accelerometer and gyroscope signals were cross-correlated with the reference force signal (FSR Force List). The lag corresponding to maximum correlation was used to temporally align each sensor channel. Accelerometer channels were aligned using a reference channel , while gyroscope channels were aligned using a gyroscope reference. Following alignment, all signals were truncated to the minimum common length to ensure synchronized time series.
\subsubsection{Artifact Correction}
IMU signals occasionally exhibited transient artifacts characterized by rapid signal collapse toward zero. These segments are approximately 0.12 seconds or less and likely caused by the vibration motor stopping then restarting. These segments were detected using a derivative-based approach: the first derivative of each signal was computed after optional median filtering, and segments where the absolute derivative exceeded one standard deviation of the derivative were flagged. Candidate segments were required to meet minimum duration criteria (0.12 seconds) and demonstrate a significant amplitude drop (drop ratio threshold: 500× reduction relative to pre-segment amplitude). Detected artifact segments were interpolated using linear interpolation between the signal values immediately preceding and following each segment, with an additional recovery extension period (0.02 seconds) appended to account for signal recovery ramps.

\subsubsection{Feature Engineering}
Two feature sets were constructed for separate modeling approaches. For absolute force prediction, following signal processing, cleaned aligned signals from all six IMU channels (three accelerometer axes: X, Y, Z; three gyroscope axes: X, Y, Z) were used as features. Additionally, magnitude features were computed as the Euclidean norm of the three-axis accelerometer and gyroscope vectors. For relative force prediction, a subset of features was selected: the percentile-scaled accelerometer X-axis signal and the percentile-scaled cleaned gyroscope Y-axis signal. Percentile scaling was performed by mapping each signal to a 0–100\% range using the 5th and 95th percentiles as anchors. Force measurements were similarly processed: absolute force was converted to pounds (conversion factor: 0.224809), while relative force was computed as percentage change using 5th and 95th percentile normalization.

\subsubsection{Model Training and Evaluation}
Ridge regression models were trained separately for absolute and relative force prediction. For the absolute force model, features were standardized using StandardScaler prior to regression. The relative force model used unscaled features. Model performance was evaluated using hold-one-out cross-validation, where each measurement session was held out as a test set while the remaining measurement sessions were used for training. This approach ensured that predictions were evaluated on completely unseen measurements, providing a robust assessment of model generalizability. Performance metrics included coefficient of determination (R²), mean absolute error (MAE), and root mean squared error (RMSE). Per-measurement error analysis was conducted to identify potential outlier measurements using a threshold of mean MAE plus two standard deviations.

% \subsection{Hand Grip Strength Estimation}
% For each smartphone, data from the force damping effect profiling study provides IMU data during a range of applied force values. To estimate force, the IMU profile is used to train a machine learning model for each phone model. The IMU data contains three axis accelerometer data and three axis gyroscope data. Each axis of the IMU data contains high frequency signals from the vibration motor as well as noise at lower frequencies. The applied force from the finger most significantly affects the amplitude of the signal oscillation. To leverage this understanding, the rolling standard deviation of each axis of the accelerometer and gyroscope data serve as features. The features most significantly correlated to the force data are included as model inputs, while other features are excluded. With these IMU data features as inputs, a regression model (multivariate linear regression model or gradient boosting) is trained using a ten percent holdout for validation. This process is repeated for each smartphone such that a distinct force estimation model exists for each smartphone model.

\begin{figure}[h!]
    \centering
    \includegraphics[width=8cm]{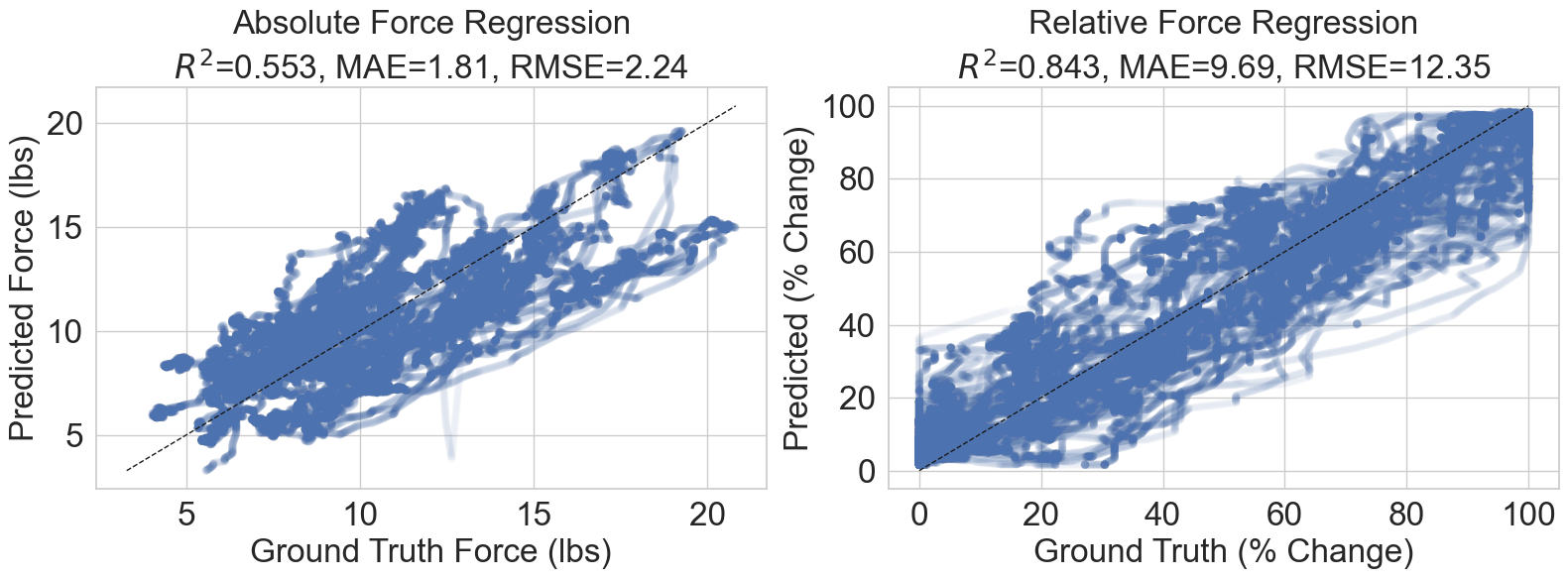}
    \caption{Regression plots of model performance for absolute applied force (left) and relative applied force (right).}
    \label{fig:example}
\end{figure}

\section{Results}

\subsection{Applied Force Measurement Validation}

\begin{figure}[ht!]
    \centering
    \includegraphics[width=8cm]{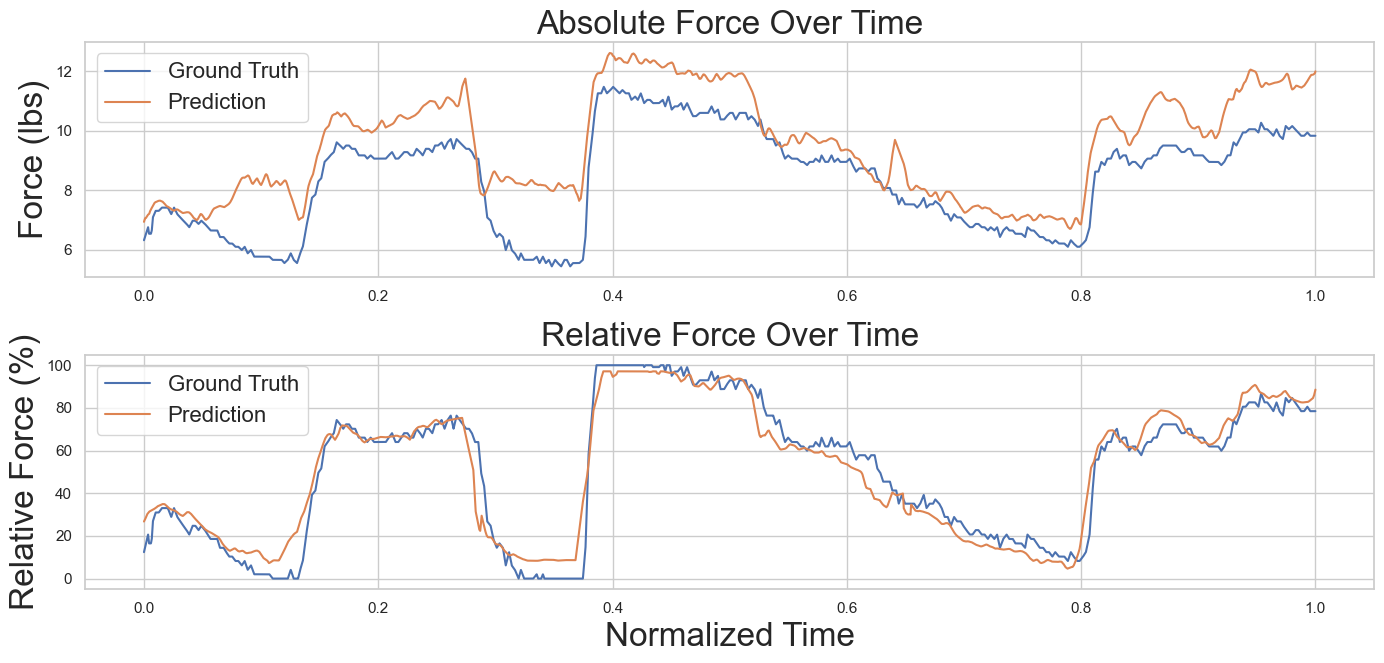}
    \caption{A measurement session sample with representative performance for absolute applied force (top) and relative applied force (bottom) estimation.}
    \label{fig:results}
\end{figure}

We evaluated the proposed vibration-based force estimation (VFE) approach using a benchtop experiment consisting of 209{,}603 six-channel IMU samples paired with synchronized ground-truth force measurements. Model performance was assessed using a 15-fold hold-one-out cross-validation scheme, where each fold corresponded to one force-profiling session. Regression results are visualized in Figure \ref{fig:results}, with an example prediction trace shown in Figure \ref{fig:example}.

Across all folds, the absolute force estimation accuracy demonstrated strong agreement with the ground truth. The mean absolute error (MAE) across folds was 1.88~lbs (SD~1.00), with the best fold achieving an MAE of 0.41~lbs and the worst fold 3.69~lbs. Similarly, the root-mean-square error (RMSE) averaged 2.10~lbs (SD~1.03), ranging from 0.51~lbs (best) to 3.90~lbs (worst). These results indicate that, when profiled for a specific smartphone model, the VFE method can reliably estimate absolute applied finger force from high-frequency vibration-modulated IMU signals.

We additionally evaluated relative force error, which quantifies estimation accuracy as a percentage of the applied force magnitude. Relative MAE averaged 10.07\% (SD~3.48\%), with fold-level performance ranging from 5.69\% to 18.73\%. Relative RMSE averaged 12.40\% (SD~4.04\%), with a best performance of 7.64\% and worst of 23.16\%. These percentage-based errors reflect natural variability in sessions with low-force ranges (where small absolute deviations inflate relative error), but remain within a range suitable for downstream tasks requiring continuous force tracking.

\section{Discussion}
The results demonstrate both significant potential and possible limitations. The absolute prediction of applied force is significantly worse than the relative force prediction. This is understandable and expected because the absolute force prediction requires a more generalized model. While the performance of the relative force estimation suggests improvements could be made to improve the absolute force prediction, the relative force metric alone has limited use cases. The absolute force metric would likely be the plausible metric for clinical and fitness applications. 

The other consideration is that the smartphone metric is a proxy for hand grip strength. The measurements are performed by squeezing the smartphone between the thumb and index finger. This metric is similar to the more common hand grip strength positioning between all 4 fingers and the palm of the hand, but further studies may be needed to assess the impact this difference may have within different application domains. The more standard hand grip positioning demonstrated that the measurement is likely feasible on a smartphone; however, the positioning of the hand and how the user applied force created greater variability within the sensor readings. For example, the user applies the majority of force with the index finger and pinky finger, the user's applied force is more akin to bending the phone around the palm rather than directly pressing the phone into the palm. These signals can look quite different. 

An additional concern that was reduced by switching to the finger pinching positioning is the potential for harmful physical effects from the measurement. Further investigation is warranted to investigate if the repeated use of this style of measurement could induce potential harmful effects like bruising, arthritis, damage to the smartphone, etc. Participants are generally able to exert far more force with full hand grip versus the finger pinch so opting for the lower applied forces reduces this risk.

\section{Conclusions}
With an MAE of less than 2lbs, this study suggests significant potential for smartphone HGS measurements. There are an abundance of clinical opportunities for this low risk, low burden HGS measurement performed with only a smartphone. The scalability of this measurement should be considered with an understanding of the phone model profiling procedure. Before a specific smartphone model can perform this measurement, the smartphone model must be profiled to train a specific force estimation model. However, profiling must only be performed once for each smartphone model, then the measurement functionality can be performed on all the hundreds of thousands of identical smartphones manufactured.

% Future studies - larger scale, detection of various medical conditions
This work provides significant opportunities for future work on applications and improvements. Investigating calibration techniques or physical modeling of the force damping system could provide methods to enable measurement functionality without specific smartphone model profiling. Additionally, the measurement accuracy could likely be improved by more complex models and larger scale data collection for smartphone profiling. On the clinical side, larger scale studies on various medical conditions that influence HGS could provide a far better assessment on potential applications and limitations of this technology.

\section{Acknowledgements}
This research was supported by the National Institutes of Health/National Institute on Aging, Trailblazer Award 5R21AG084975.

\section{References}
\printbibliography

@article{kirk_sarcopenia_2021,
	title = {Sarcopenia Definitions and Outcomes Consortium ({SDOC}) Criteria are Strongly Associated With Malnutrition, Depression, Falls, and Fractures in High-Risk Older Persons},
	volume = {22},
	issn = {15258610},
	url = {https://linkinghub.elsevier.com/retrieve/pii/S1525861020305727},
	doi = {10.1016/j.jamda.2020.06.050},
	pages = {741--745},
	number = {4},
	journaltitle = {Journal of the American Medical Directors Association},
	shortjournal = {Journal of the American Medical Directors Association},
	author = {Kirk, Ben and Zanker, Jesse and Bani Hassan, Ebrahim and Bird, Stefanie and Brennan-Olsen, Sharon and Duque, Gustavo},
	urldate = {2025-03-27},
	date = {2021-04},
	langid = {english},
}

@article{mayhew_sarcopenia_2023,
	title = {Sarcopenia Definition and Outcomes Consortium 2020 Definition: Association and Discriminatory Accuracy of Sarcopenia With Disability in the Canadian Longitudinal Study on Aging},
	volume = {78},
	rights = {https://academic.oup.com/pages/standard-publication-reuse-rights},
	issn = {1079-5006, 1758-535X},
	url = {https://academic.oup.com/biomedgerontology/article/78/9/1597/7179735},
	doi = {10.1093/gerona/glad131},
	shorttitle = {Sarcopenia Definition and Outcomes Consortium 2020 Definition},
	pages = {1597--1603},
	number = {9},
	journaltitle = {The Journals of Gerontology: Series A},
	author = {Mayhew, Alexandra J and Sohel, Nazmul and Beauchamp, Marla K and Phillips, Stuart and Raina, Parminder},
	editor = {Lipsitz, Lewis A},
	urldate = {2025-03-27},
	date = {2023-08-27},
	langid = {english},
}

@article{cruz-jentoft_sarcopenia_2010,
	title = {Sarcopenia: European consensus on definition and diagnosis},
	volume = {39},
	rights = {https://creativecommons.org/licenses/by-nc/2.5/},
	issn = {1468-2834, 0002-0729},
	url = {https://academic.oup.com/ageing/article/39/4/412/8732},
	doi = {10.1093/ageing/afq034},
	shorttitle = {Sarcopenia},
	pages = {412--423},
	number = {4},
	journaltitle = {Age and Ageing},
	author = {Cruz-Jentoft, Alfonso J. and Baeyens, Jean Pierre and Bauer, Jürgen M. and Boirie, Yves and Cederholm, Tommy and Landi, Francesco and Martin, Finbarr C. and Michel, Jean-Pierre and Rolland, Yves and Schneider, Stéphane M. and Topinková, Eva and Vandewoude, Maurits and Zamboni, Mauro},
	urldate = {2025-03-27},
	date = {2010-07-01},
	langid = {english},
}

@article{rijk_prognostic_2016,
	title = {Prognostic value of handgrip strength in people aged 60 years and older: A systematic review and meta‐analysis},
	volume = {16},
	rights = {http://onlinelibrary.wiley.com/{termsAndConditions}\#vor},
	issn = {1444-1586, 1447-0594},
	url = {https://onlinelibrary.wiley.com/doi/10.1111/ggi.12508},
	doi = {10.1111/ggi.12508},
	shorttitle = {Prognostic value of handgrip strength in people aged 60 years and older},
	pages = {5--20},
	number = {1},
	journaltitle = {Geriatrics \& Gerontology International},
	shortjournal = {Geriatrics Gerontology Int},
	author = {Rijk, Joke M and Roos, Paul Rkm and Deckx, Laura and Van Den Akker, Marjan and Buntinx, Frank},
	urldate = {2025-04-02},
	date = {2016-01},
	langid = {english},
}

@article{sousa-santos_differences_2017,
	title = {Differences in handgrip strength protocols to identify sarcopenia and frailty - a systematic review},
	volume = {17},
	issn = {1471-2318},
	url = {http://bmcgeriatr.biomedcentral.com/articles/10.1186/s12877-017-0625-y},
	doi = {10.1186/s12877-017-0625-y},
	pages = {238},
	number = {1},
	journaltitle = {{BMC} Geriatrics},
	shortjournal = {{BMC} Geriatr},
	author = {Sousa-Santos, A. R. and Amaral, T. F.},
	urldate = {2025-04-02},
	date = {2017-12},
	langid = {english},
}

@article{myles_how_2024,
	title = {The how and why of handgrip strength assessment},
	volume = {87},
	issn = {0308-0226, 1477-6006},
	url = {https://journals.sagepub.com/doi/10.1177/03080226231208409},
	doi = {10.1177/03080226231208409},
	pages = {321--328},
	number = {5},
	journaltitle = {British Journal of Occupational Therapy},
	shortjournal = {British Journal of Occupational Therapy},
	author = {Myles, Louise and Massy-Westropp, Nicola and Barnett, Fiona},
	urldate = {2025-04-02},
	date = {2024-05},
	langid = {english},
}

@article{barry_oscillometric_2024,
	title = {Oscillometric blood pressure measurements on smartphones using vibrometric force estimation},
	volume = {14},
	issn = {2045-2322},
	url = {https://www.nature.com/articles/s41598-024-75025-9},
	doi = {10.1038/s41598-024-75025-9},
	pages = {26206},
	number = {1},
	journaltitle = {Scientific Reports},
	shortjournal = {Sci Rep},
	author = {Barry, Colin and Xuan, Yinan and Fascetti, Ava and Moore, Alison and Wang, Edward Jay},
	urldate = {2025-12-01},
	date = {2024-10-31},
	langid = {english},
}

@inproceedings{zhang_vibroscale_2020,
	location = {Virtual Event Mexico},
	title = {{VibroScale}: turning your smartphone into a weighing scale},
	isbn = {978-1-4503-8076-8},
	url = {https://dl.acm.org/doi/10.1145/3410530.3414397},
	doi = {10.1145/3410530.3414397},
	shorttitle = {{VibroScale}},
	eventtitle = {{UbiComp}/{ISWC} '20: 2020 {ACM} International Joint Conference on Pervasive and Ubiquitous Computing and 2020 {ACM} International Symposium on Wearable Computers},
	pages = {176--179},
	booktitle = {Adjunct Proceedings of the 2020 {ACM} International Joint Conference on Pervasive and Ubiquitous Computing and Proceedings of the 2020 {ACM} International Symposium on Wearable Computers},
	publisher = {{ACM}},
	author = {Zhang, Shibo and Xu, Qiuyang and Sen, Sougata and Alshurafa, Nabil},
	urldate = {2025-12-01},
	date = {2020-09-10},
	langid = {english},
}

@inproceedings{goel_gripsense_2012,
	location = {Cambridge Massachusetts {USA}},
	title = {{GripSense}: using built-in sensors to detect hand posture and pressure on commodity mobile phones},
	isbn = {978-1-4503-1580-7},
	url = {https://dl.acm.org/doi/10.1145/2380116.2380184},
	doi = {10.1145/2380116.2380184},
	shorttitle = {{GripSense}},
	eventtitle = {{UIST} '12: The 25th Annual {ACM} Symposium on User Interface Software and Technology},
	pages = {545--554},
	booktitle = {Proceedings of the 25th annual {ACM} symposium on User interface software and technology},
	publisher = {{ACM}},
	author = {Goel, Mayank and Wobbrock, Jacob and Patel, Shwetak},
	urldate = {2025-12-01},
	date = {2012-10-07},
	langid = {english},
}

@article{huang_association_2022,
	title = {Association between grip strength and cognitive impairment in older American adults},
	volume = {15},
	issn = {1662-5099},
	url = {https://www.frontiersin.org/articles/10.3389/fnmol.2022.973700/full},
	doi = {10.3389/fnmol.2022.973700},
	pages = {973700},
	journaltitle = {Frontiers in Molecular Neuroscience},
	shortjournal = {Front. Mol. Neurosci.},
	author = {Huang, Jian and Wang, Xinping and Zhu, Hao and Huang, Dong and Li, Weiwang and Wang, Jing and Liu, Zhirong},
	urldate = {2023-03-29},
	date = {2022-11-30},
}

@article{rantanen_midlife_1999,
	title = {Midlife Hand Grip Strength as a Predictor of Old Age Disability},
	volume = {281},
	issn = {0098-7484},
	url = {http://jama.jamanetwork.com/article.aspx?doi=10.1001/jama.281.6.558},
	doi = {10.1001/jama.281.6.558},
	pages = {558},
	number = {6},
	journaltitle = {{JAMA}},
	shortjournal = {{JAMA}},
	author = {Rantanen, Taina},
	urldate = {2023-03-28},
	date = {1999-02-10},
	langid = {english},
}

@article{mcgrath_handgrip_2018,
	title = {Handgrip Strength and Health in Aging Adults},
	volume = {48},
	issn = {0112-1642, 1179-2035},
	url = {http://link.springer.com/10.1007/s40279-018-0952-y},
	doi = {10.1007/s40279-018-0952-y},
	pages = {1993--2000},
	number = {9},
	journaltitle = {Sports Medicine},
	shortjournal = {Sports Med},
	author = {{McGrath}, Ryan P. and Kraemer, William J. and Snih, Soham Al and Peterson, Mark D.},
	urldate = {2023-03-28},
	date = {2018-09},
	langid = {english},
}

@article{bohannon_grip_2019,
	title = {Grip Strength: An Indispensable Biomarker For Older Adults},
	volume = {Volume 14},
	issn = {1178-1998},
	url = {https://www.dovepress.com/grip-strength-an-indispensable-biomarker-for-older-adults-peer-reviewed-article-CIA},
	doi = {10.2147/CIA.S194543},
	shorttitle = {Grip Strength},
	pages = {1681--1691},
	journaltitle = {Clinical Interventions in Aging},
	shortjournal = {{CIA}},
	author = {Bohannon, Richard W},
	urldate = {2023-03-28},
	date = {2019-10},
	langid = {english},
}

@article{griffith_delayed_2005,
	title = {Delayed recovery of hand grip strength predicts postoperative morbidity following major vascular surgery},
	volume = {76},
	issn = {0007-1323, 1365-2168},
	url = {https://academic.oup.com/bjs/article/76/7/704/6172173},
	doi = {10.1002/bjs.1800760717},
	pages = {704--705},
	number = {7},
	journaltitle = {British Journal of Surgery},
	author = {Griffith, C D M and Whyman, M and Bassey, E J and Hopkinson, B R and Makin, G S},
	urldate = {2022-12-13},
	date = {2005-12-06},
	langid = {english},
}

@article{song_causal_2022,
	title = {Causal associations of hand grip strength with bone mineral density and fracture risk: A mendelian randomization study},
	volume = {13},
	issn = {1664-2392},
	url = {https://www.frontiersin.org/articles/10.3389/fendo.2022.1020750/full},
	doi = {10.3389/fendo.2022.1020750},
	shorttitle = {Causal associations of hand grip strength with bone mineral density and fracture risk},
	pages = {1020750},
	journaltitle = {Frontiers in Endocrinology},
	shortjournal = {Front. Endocrinol.},
	author = {Song, Jidong and Liu, Tun and Zhao, Jiaxin and Wang, Siyuan and Dang, Xiaoqian and Wang, Wei},
	urldate = {2024-12-17},
	date = {2022-12-12},
}

@article{us_preventive_services_task_force_screening_2025,
	title = {Screening for Osteoporosis to Prevent Fractures: {US} Preventive Services Task Force Recommendation Statement},
	volume = {333},
	issn = {0098-7484},
	url = {https://jamanetwork.com/journals/jama/fullarticle/2829238},
	doi = {10.1001/jama.2024.27154},
	shorttitle = {Screening for Osteoporosis to Prevent Fractures},
	pages = {498},
	number = {6},
	journaltitle = {{JAMA}},
	shortjournal = {{JAMA}},
	author = {{US Preventive Services Task Force} and Nicholson, Wanda K. and Silverstein, Michael and Wong, John B. and Chelmow, David and Coker, Tumaini Rucker and Davis, Esa M. and Jaén, Carlos Roberto and Krousel-Wood, Marie and Lee, Sei and Li, Li and Mangione, Carol M. and Ogedegbe, Gbenga and Rao, Goutham and Ruiz, John M. and Stevermer, James and Tsevat, Joel and Underwood, Sandra Millon and Wiehe, Sarah},
	urldate = {2025-03-27},
	date = {2025-02-11},
	langid = {english},
}

@article{meza-valderrama_sarcopenia_2021,
	title = {Sarcopenia, Malnutrition, and Cachexia: Adapting Definitions and Terminology of Nutritional Disorders in Older People with Cancer},
	volume = {13},
	rights = {https://creativecommons.org/licenses/by/4.0/},
	issn = {2072-6643},
	url = {https://www.mdpi.com/2072-6643/13/3/761},
	doi = {10.3390/nu13030761},
	shorttitle = {Sarcopenia, Malnutrition, and Cachexia},
	pages = {761},
	number = {3},
	journaltitle = {Nutrients},
	shortjournal = {Nutrients},
	author = {Meza-Valderrama, Delky and Marco, Ester and Dávalos-Yerovi, Vanesa and Muns, Maria Dolors and Tejero-Sánchez, Marta and Duarte, Esther and Sánchez-Rodríguez, Dolores},
	urldate = {2025-09-29},
	date = {2021-02-26},
	langid = {english},
}

@article{chai_comparison_2024,
	title = {Comparison of grip strength measurements for predicting all-cause mortality among adults aged 20+ years from the {NHANES} 2011–2014},
	volume = {14},
	issn = {2045-2322},
	url = {https://www.nature.com/articles/s41598-024-80487-y},
	doi = {10.1038/s41598-024-80487-y},
	pages = {29245},
	number = {1},
	journaltitle = {Scientific Reports},
	shortjournal = {Sci Rep},
	author = {Chai, Lirong and Zhang, Dongfeng and Fan, Junning},
	urldate = {2025-12-01},
	date = {2024-11-25},
	langid = {english},
}

\end{document}